\documentclass[conference]{IEEEtran}

\usepackage[hyphens]{url}
\usepackage[hidelinks]{hyperref}
\usepackage[T1]{fontenc}
\usepackage[cmintegrals]{newtxmath}
\usepackage{cite}
\usepackage{bm}
\usepackage{changepage}
\usepackage{subfigure}
\usepackage{mathtools}
\usepackage[boxruled]{algorithm2e}
\usepackage{blindtext, graphicx}
\usepackage{relsize}
\usepackage{amssymb}
\usepackage{array}
\usepackage{longtable}
\usepackage[long,nocomma]{optidef}
\usepackage{framed}
\usepackage[table,xcdraw]{xcolor}

\newcommand{\bwrh}{{\sf{BWRH}}}

\newcommand{\bwrhf}{{\sf{BWRHF}}}

\begin{document}

\title{Optimizing Inter-Datacenter Tail Flow Completion Times using Best Worst-case Routing}

\author{
\IEEEauthorblockN{Max Noormohammadpour, Ajitesh Srivastava, Cauligi S. Raghavendra}
\IEEEauthorblockA{Ming Hsieh Department of Electrical Engineering, University of Southern California (USC)\\\{noormoha, ajiteshs, raghu\}@usc.edu}
}

\maketitle

\thispagestyle{plain}
\pagestyle{plain}

\begin{abstract}
Flow routing over inter-datacenter networks is a well-known problem where the network assigns a path to a newly arriving flow potentially according to the network conditions and the properties of the new flow. An essential system-wide performance metric for a routing algorithm is the flow completion times, which affect the performance of applications running across multiple datacenters. Current static and dynamic routing approaches do not take advantage of flow size information in routing, which is practical in a controlled environment such as inter-datacenter networks that are managed by the datacenter operators. In this paper, we discuss Best Worst-case Routing (BWR), which aims at optimizing the tail completion times of long-running flows over inter-datacenter networks with non-uniform link capacities. Since finding the path with the best worst-case completion time for a new flow is NP-Hard, we investigate two heuristics, \bwrh~and \bwrhf, which use two different upper bounds on the worst-case completion times for routing. We evaluate \bwrh~and \bwrhf~against several real WAN topologies and multiple traffic patterns. Although \bwrh~better models the BWR problem, \bwrh~and \bwrhf~show negligible difference across various system-wide performance metrics, while \bwrhf~being significantly faster. Furthermore, we show that compared to other popular routing heuristics, \bwrhf~can reduce the mean and tail flow completion times by over $1.5\times$ and $2\times$, respectively.

\end{abstract}

\begin{IEEEkeywords}
Flow Routing, Flow Completion Times, Path Selection, Software Defined Networking.
\end{IEEEkeywords}

\IEEEpeerreviewmaketitle

\section{Introduction}
We revisit the well-known flow routing problem over the inter-datacenter networks where one operator manages the network routing as well as the datacenters. Examples of such networks are Google B4 \cite{b4, b4andafter}, Facebook Express Backbone \cite{facebook-express-backbone}, and Microsoft Global WAN \cite{swan-backbone}. Traffic forwarding over these networks is controlled in a logically centralized fashion which brings about new opportunities. That is, routing algorithms can use a global view of network status and end-point demands to optimize the routing of inter-datacenter traffic.


A growing fraction of traffic over the inter-datacenter networks is generated by cross-datacenter flows that replicate content and data across many datacenters \cite{facebook-express-backbone}. These flows are mostly throughput-oriented, carry large volumes of data, and are resilient to initial routing and scheduling latency. Optimizing the completion times of cross-datacenter flows, which is the focus of this paper, can improve the performance of geographically distributed datacenter applications and increase the overall network utility.


An inter-datacenter flow can be indicated with a source datacenter, a receiver datacenter, and the volume of data to be moved. We consider an online scenario without knowledge of future flow arrivals. Upon arrival, the routing algorithm with a global network view will select a path for the new flow. Similar to the standard routing techniques, once a path is selected for a flow, it cannot be changed unless there is a network failure. The objective is to minimize the tail completion time of flows over a given period where a flow's completion time is defined as the time since its arrival to its completion. Since the tail-optimal flow completion times can only be computed given the knowledge of future flow arrivals, we apply an online-greedy approach to minimizing the tail flow completion times named Best Worst-case Routing (BWR).\footnote{An earlier version of BWR was proposed and analyzed in \cite{bwr}. In this paper, we have extended BWR to networks with non-uniform link capacities and have also proposed and evaluated an additional heuristic with guaranteed polynomial running time.}

We briefly go over current online path selection techniques. Multiple static and dynamic routing techniques have been proposed for this purpose, which consider link capacity information, the number of hops, link propagation latency, and instantaneous link utilization for routing \cite{tvlakshman, routing-metric}. For example, the shortest widest path approach \cite{tvlakshman} selects the path that offers the maximum available bandwidth at the current time with the minimum number of hops from the source to the destination. Also, the min-max utilization approach selects the path with the minimum number of hops over which the maximum utilization is minimum from the source to the destination. Besides, a popular static routing approach is to assign to every edge a cost equal to their inverse capacity and then select one shortest path from the source to the destination \cite{routing-metric}. We will consider these approaches later in the evaluations.

The BWR technique presented in this work takes advantage of the remaining flow size information for long-running flows to further improve on the current online and dynamic routing approaches. In a network that is managed in a logically centralized fashion, such information is either known by the traffic scheduler (e.g., \cite{tempus, amoeba, owan}) or can be obtained from the end-points periodically. In summary, our paper makes the following contributions:

\begin{itemize}
    \setlength{\itemsep}{0.5em}
    \item We propose the Best Worst-case Routing (BWR) with aims to minimize the worst-case completion time of every incoming flow given the network topology, the currently ongoing flows' paths, and their remaining volume of data. We make no assumptions on the future flow arrivals and impose no constraints on the traffic scheduling policy.
    
    \item Since BWR is NP-Hard \cite{bwr}, we propose two heuristics that aim to select a path that minimizes an upper bound on the worst-case completion time of the arriving flow. We discuss why computing the exact worst-case per path can be computationally intensive as it may require building complex dependency graphs.
    
    \item We run extensive simulations to compare the performance of \bwrh~and \bwrhf~with that of popular static and dynamic routing approaches currently used. We first find that \bwrh~and \bwrhf~offer almost identical performance across various topologies and traffic patterns. Next, we show that \bwrhf~can improve the mean and tail completion times by over $1.5\times$ and $2\times$, respectively, given various flow size distributions and scheduling policies.
\end{itemize}

\section{System Model} \label{model}
Let us consider an inter-datacenter network with non-uniform link capacity distribution across the edges. We assume an online scenario without the knowledge of future flow arrivals. Every flow is assigned a fixed path when it arrives which is computed using the information available at a logically centralized network controller on the network topology and the other currently ongoing flows. This assumption is based on many recent related works on management of inter-datacenter WAN \cite{b4, facebook-express-backbone, swan-backbone}. We also assume the availability of flow size information for the new flow and the remaining flow size for ongoing flows. Please find the definition of variables used in this paper in Table \ref{table_var}.

\begin{table}[b]
    \small
    \begin{center}
    \caption{Definition of Variables} \label{table_var}
    \begin{tabular}{ |p{1.2cm}|p{6.5cm}| }
        \hline
        \textbf{Variable} & \textbf{Definition} \\
        \hline
        \hline
        $F_i$ & A data flow \\
        \hline
        $s_i$ & Source node of $F_i$ \\
        \hline
        $t_i$ & Destination node of $F_i$ \\
        \hline
        $\alpha_i$ & The arrival time of $F_i$ \\
        \hline
        $\beta_i$ & Worst-case finish time of $F_i$ \\
        \hline
        $\gamma_i$ & Worst-case completion time of $F_i$, i.e., $\gamma_i = \beta_i - \alpha_i$ \\
        \hline
        $\mathcal{V}_i$ & Total volume of $F_i$ in bytes \\
        \hline
        $\mathcal{V}^r_i$ & Remaining volume of $F_i$ in bytes, i.e., $\mathcal{V}^r_i \le \mathcal{V}_i$ \\
        \hline
        $G(V,E)$ & The inter-datacenter network graph \\
        \hline
        $e$ & An edge where $e \in E$ given $G(V,E)$ \\
        \hline
        $C_e$ & Capacity of edge $e$ in bytes per second \\
        \hline
        $P$ & A path on the inter-datacenter network \\
        \hline
        $P_i$ & Path selected for $F_i$ \\
        \hline
        $\gamma^P_i$ & Worst-case completion time of $F_i$ on path $P$ \\
        \hline
        $E_{P_i}$ & Edges of path $P_i$ \\
        \hline
        $\mathcal{F}_e$ & All flows whose paths goes over $e$ \\
        \hline
    \end{tabular}
    \end{center}
\end{table}

We focus on the routing of long flows for which the additional latency due to centralized control is acceptable. Also, by aiming at sufficiently large flows, the queuing and propagation latency can be negligible and so can be omitted from the calculations. The routing for short flows can be done in a distributed fashion without the involvement of the logically centralized controller, e.g., short flows can be routed on the paths with minimum propagation latency. This approach will make the system more scalable as the number of large flows is usually a small fraction of the total number of flows although they carry significant volumes of traffic \cite{social_inside, facebook-express-backbone}. \footnote{A flow is large if its completion time is orders of magnitude larger than the propagation latency or if it is orders of magnitude larger than the median flow size.}

Depending on how network queues prioritize the transmission of data packets and how the end-points transmit traffic, a variety of network scheduling policies can be realized. We consider the three popular scheduling policies of First Come First Serve (FCFS), Shortest Remaining Processing Time (SRPT), and fair sharing based on max min fairness \cite{max-min-fairness}. In evaluations, we will focus mainly on the fair sharing policy as that is the ideal outcome when multiple TCP flows share network links.

Finally, we assume that the network controller is capable of dynamically updating the network forwarding state to install custom routes for the long incoming flows. Such forwarding is achievable with the application of Software Defined Networking (SDN) \cite{sdn} which is supported by several current inter-datacenter networks \cite{b4, swan-backbone, facebook-express-backbone}. This goal is obtainable over MPLS networks as well via techniques such as Segment Routing (SR) \cite{filsfils2018segment}.

\section{Best Worst-case Routing (BWR)}
BWR aims to minimize the worst-case (i.e., tail) completion time of an incoming flow without any knowledge of future flow arrivals. The worst-case completion time of a flow is independent of the network scheduling policy and is only a function of network topology, links' capacities, and the remaining volumes of traffic for the current flows. The BWR problem can be stated as follows (variables in Table \ref{table_var}):

\vspace{0.5em}
\noindent\textbf{Best Worst-case Routing Problem:} \textit{Given an inter-datacenter wide area network $G(V,E)$, capacity $C_e$ for $e \in E$, and current flows $F_i, 1 \le i \le N$, we want to assign a path $P_{N+1}$ to the new flow $F_{N+1}$ so that $\max(\gamma_{N+1})$ is minimized.}

\vspace{0.5em}
This is a greedy online approach to optimizing the tail completion times of flows. This problem is highly complex for two reasons. First, in a general setting, computing the exact worst-case completion time of $F_{N+1}$ on a given path $P$ may be computationally intensive due to the interdependence of current flows.\footnote{Please see the appendix for an example and discussion.} Next, even assuming that there is an efficient algorithm to compute the worst-case completion time of $F_{N+1}$ on any given path $P$, finding the path $P_{N+1}$ is NP-Hard \cite{bwr}.

\vspace{0.5em}
\noindent\textbf{Our Approach:} Given that BWR is NP-Hard, we propose two heuristics in the following sections for finding an approximate solution. Instead of computing the exact worst-case completion time of $F_{N+1}$ for a given path $P$ (i.e., exact $\gamma^P_{N+1}$), we use two different upper bounds. Depending on the upper bound used, the heuristics developed will have different properties.

\section{\bwrh}
We compute an upper bound for the worst-case completion time of a new flow given the following assumptions. First, we assume no knowledge of future arrivals, and so only the current flows are considered. Second, the worst-case occurs when all the current flows that have a common edge with a candidate path $P$ are sent before $F_{N+1}$ and that their bottleneck is the edge with minimum capacity that is in common with $P$. Next, the worst-case occurs when all the current flows that have a common edge with $P$ are transmitted sequentially, and so the time to complete their remaining volumes is accumulated. Finally, after all the current flows that have a common edge with $P$ are finished, it will take $F_{N+1}$ equal to its volume divide by the bottleneck capacity on $P$ to complete. Therefore, given the ongoing flows $F_i, 1 \le i \le N$, we approximate $\gamma^P_{N+1}$ for any given path $P$ using the aforementioned upper bound:

\begin{align}
    \gamma^P_{N+1} \approx \sum_{\substack{1 \le i \le N\\E_P \cap E_{P_i} \neq \emptyset}} \frac{\mathcal{V}^r_{i}}{\min_{e \in \{E_P \cap E_{P_i}\}}(C_e)} + \frac{\mathcal{V}_{{N+1}}}{\min_{e \in E_P}(C_e)} \label{cost_bwrh}
\end{align}

A straightforward approach to computing the path with the best worst-case completion time for $F_{N+1}$ is using exhaustive search. However, since the number of paths between two end-points grows exponentially with the network size, exhaustive search is generally inefficient and slow. In Algorithm \ref{bwrh}, we have augmented exhaustive search with a straightforward termination condition that speeds up the process significantly for the majority of flows.

\SetAlgoVlined
\begin{algorithm}[t]
\caption{\bwrh} \label{bwrh}
{
\SetKw{KwBy}{by}
\SetKwProg{FindPath}{FindPath}{}{}

\vspace{0.4em}
\KwIn{$F_{N+1}$, $G(V,E)$, $P_i$, and $\mathcal{V}^r_i,1 \le i \le N$}

\vspace{0.4em}
\KwOut{$P_{N+1}$}

\hrulefill

\vspace{0.4em}
$K \gets$ \#hops on the minimum hop path from $s_{N+1}$ to $t_{N+1}$\;

\vspace{0.4em}
$P^K_{min} \gets$ Find a path $P$ with at most $K$ hops from $s_{N+1}$ to $t_{N+1}$ for which $\gamma^P_{N+1}$ is the minimum possible by scanning all such paths\;

\vspace{0.4em}
\Repeat{$\gamma^{P^K_{min}}_{N+1} < \gamma^{P^{K-1}_{min}}_{N+1}$}{
    \vspace{0.4em}
    $K \gets K+1$\;
    
    \vspace{0.4em}
    Compute $P^K_{min}$\;
    
    \vspace{0.4em}
}

\vspace{0.6em}
$P_{N+1} \gets P^K_{min}$\;
}
\end{algorithm}

Algorithm \ref{bwrh} finds a path $P_{N+1}$ for $F_{N+1}$ using the approximation in Eq. \ref{cost_bwrh}. At every iteration, the algorithm finds the path $P^K_{min}$ with at most $K$ hops from $s_{N+1}$ to $t_{N+1}$ with the minimum value of $\gamma^{P^K_{min}}_{N+1}$ by finding and examining all such paths. Assuming that $K_0$ is the number of hops on the minimum hop path from $s_{N+1}$ to $t_{N+1}$, the algorithm finds $P^{K_0}_{min}$ and $\gamma^{P^{K_0}_{min}}_{N+1}$ where $K = K_0$. It then increases the number of maximum hops allowed by one, i.e., $K = K+1$, extending the search space to more paths. This process continues until there is no improvement in the worst-case completion time of the best path, i.e., the path for which the worst-case completion time is minimum while increasing $K$.

The termination condition in \bwrh~favors shorter paths. If the best path is much longer than the minimum hop path, it is possible that the algorithm terminates before it is found. However, it is unlikely, in general, for the best path to be long as having more edges increases the likelihood of having common edges with more current flows, which increases the worst-case completion time on a path.

\section{\bwrhf}
The worst-case running time of Algorithm \ref{bwrh} is exponential as it might search all existing paths from $s_{N+1}$ to $t_{N+1}$. To improve, we propose a new heuristic that runs in guaranteed polynomial time. We do that by using a new upper bound on the worst-case completion time of a new flow which is based on the following inequality:

\begin{align}
    \frac{1}{\min(x_1,\dots,x_n)} \le \frac{1}{x_1}+\dots+\frac{1}{x_n} \label{inequality}
\end{align}

{\noindent}Where $x_i$, $1 \le i \le n$, are positive real numbers. Accordingly, we can arrive at the following upper bound from the one in Eq. \ref{cost_bwrh}. That is, given the ongoing flows $F_i, 1 \le i \le N$, we approximate $\gamma^P_{N+1}$ for any given path $P$ with:

\begin{align}
    \gamma^P_{N+1} \approx \sum_{e \in E_P} \Big(\frac{\sum_{F_i \in \mathcal{F}_e}\mathcal{V}^r_{i}+\mathcal{V}_{N+1}}{C_e}\Big) \label{cost_bwrhf}
\end{align}

Algorithm \ref{bwrhf} shows how \bwrhf~finds a path $P_{N+1}$ for $F_{N+1}$ using this new upper bound. In contrast with the upper bound used in \bwrh, the one used in Algorithm \ref{bwrhf} is edge-decomposable. Therefore, \bwrhf~can be implemented using any standard shortest path algorithm. This algorithm offers guaranteed polynomial running time, which is a highly desired property for online flow routing.

\SetAlgoVlined
\begin{algorithm}[t]
\caption{\bwrhf} \label{bwrhf}
{
\SetKw{KwBy}{by}
\SetKwProg{FindPath}{FindPath}{}{}

\vspace{0.4em}
\KwIn{$F_{N+1}$, $G(V,E)$, $P_i$, and $\mathcal{V}^r_i,1 \le i \le N$}

\vspace{0.4em}
\KwOut{$P_{N+1}$}

\hrulefill

\vspace{0.4em}
To every edge $e \in E$, assign the weight $\frac{\sum_{F_i \in \mathcal{F}_e} \mathcal{V}^r_i + \mathcal{V}_{N+1}}{C_{e}}$\;

\vspace{0.4em}
$P_{N+1} \gets$ A minimum weight path from $s_{N+1}$ to $t_{N+1}$ found by a standard shortest path algorithm (e.g., Dijkstra's algorithm)\;
}
\end{algorithm}

\begin{figure}[t]
    \centering
    \includegraphics[width=\columnwidth]{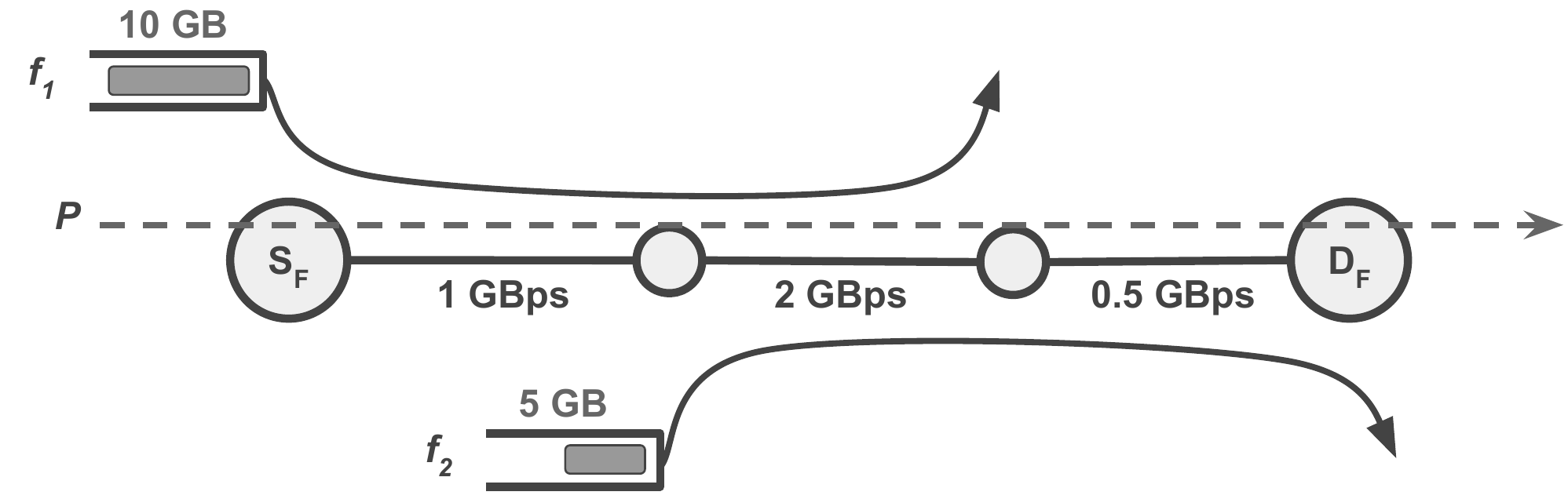}
    \caption{Example used to demonstrate the two upper bounds for $\gamma^P_3$} \label{fig:example}
\end{figure}

\begin{figure*}
    \centering
    \hspace{-0.2em}\subfigure[Average performance gap between \bwrhf~and \bwrh~over five WAN topologies]{\label{fig:exp_bwrhf_bwrh_diff} 
        \includegraphics[width=0.65\textwidth]{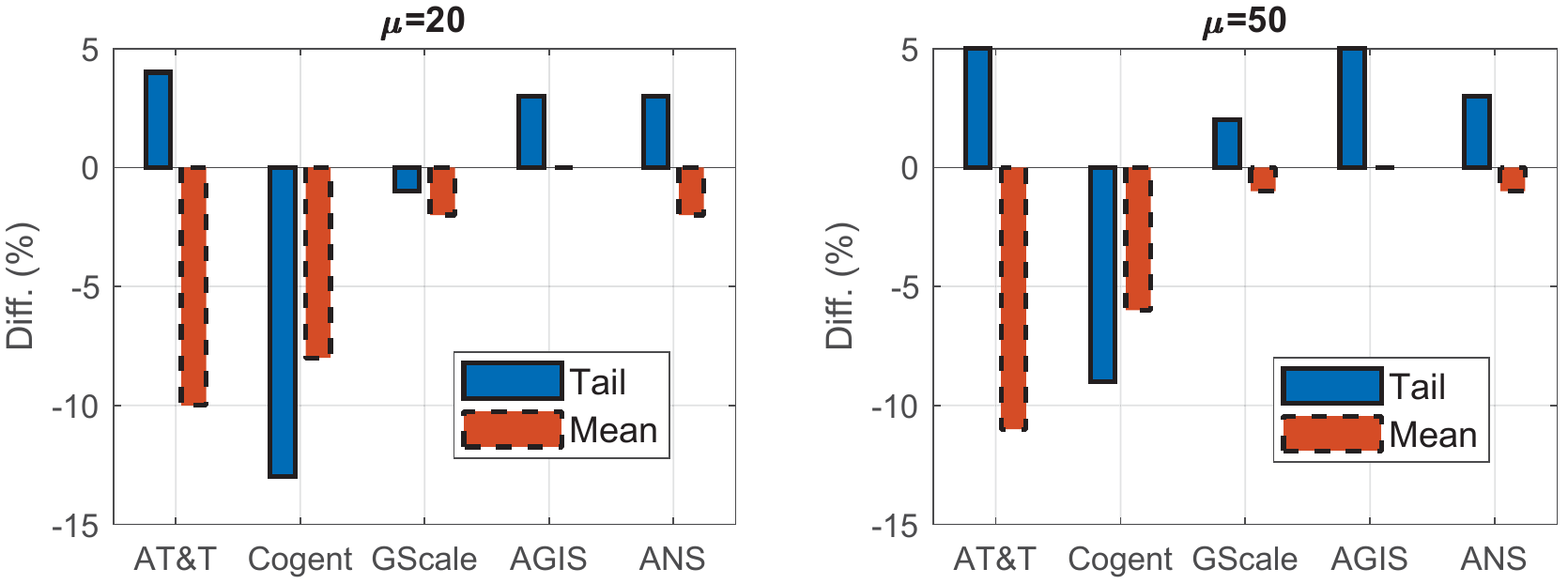}}\qquad\qquad%
    \subfigure[Worst-case running times]{\label{fig:exp_bwrhf_bwrh_running_time}
        \includegraphics[width=0.255\textwidth]{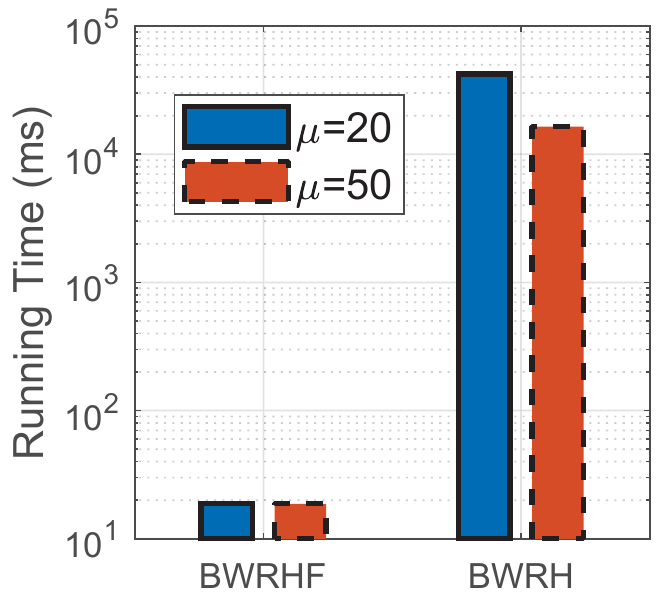}}
    \caption{Comparison of flow completion times and running time for the two heuristics of \bwrhf~and \bwrh~over three different traffic patterns (i.e., light-tailed, heavy-tailed, and cache-follower) and three scheduling policies (i.e., FCFS, SRPT, and fair sharing).}\label{fig:exp_bwrhf_bwrh}
\end{figure*}

\section{An Example}
Consider the scenario shown in Figure \ref{fig:example}. A new flow $F_3$ with a volume of 8 GB has arrived for which we are considering the candidate path $P$. We will compute the two upper bounds on $\gamma^P_3$ as discussed in the two previous sections. The approximation in Eq. \ref{cost_bwrh} is computed as follows:
\begin{align}
    \gamma^P_3 \approx \frac{10}{\min(1,2)} + \frac{5}{\min(2,0.5)} + \frac{8}{\min(1,2,0.5)} = 36
\end{align}

The approximation of Eq. \ref{cost_bwrhf} can be computed as follows:
\begin{align}
    \gamma^P_3 \approx \frac{10+8}{1} + \frac{10+5+8}{2} + \frac{5+8}{0.5} = 55.5
\end{align}

The second upper bound will always be greater than the first one as shown in Eq. \ref{inequality}.

\begin{figure*}[p]
    \centering
    \subfigure[By scheduling policy given AT\&T North America topology with randomly generated variable link capacities]{\includegraphics[width=0.8\textwidth]{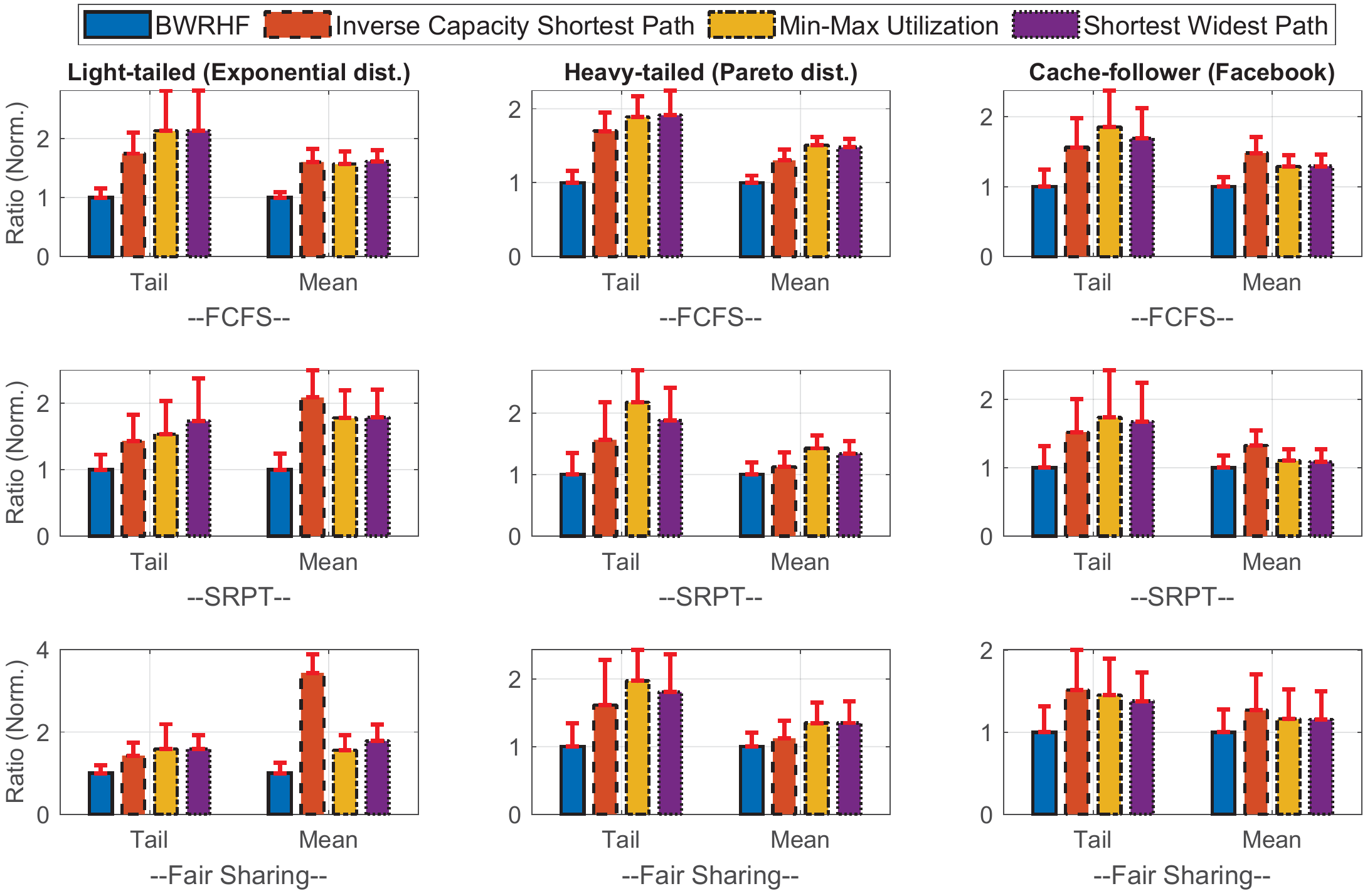} \label{fig:exp_att_policy}}\\
    \subfigure[By network topology given the fair sharing policy based on max-min fairness]{\includegraphics[width=0.8\textwidth]{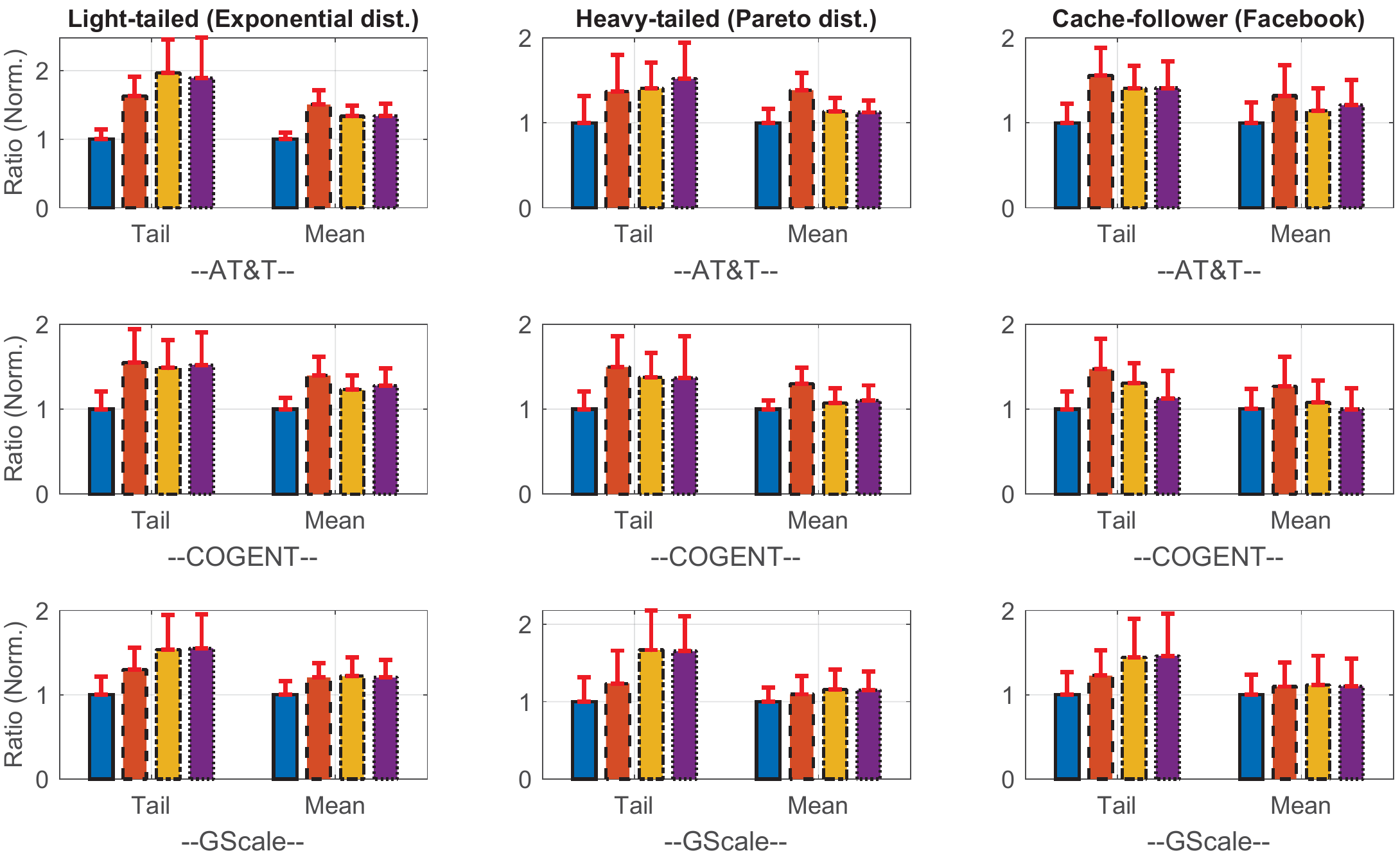} \label{fig:exp_mmf_topology}}
    \caption{Flow completion times reported for 500 flow arrivals given $\lambda=1$ and $\mu=50$ for various traffic patterns. The capacity of every link was randomly generated using a uniform distribution from 0.2 to 1, and each scenario was repeated 20 times to consider many capacity distribution possibilities. The results were averaged across the 20 runs, and the error bars show the standard deviation over the runs. The Facebook patterns were inferred from the CDF curves reported in \cite{social_inside}.}
\end{figure*}

\section{Evaluations}
We considered both synthetic flow size distributions of light-tailed (Exponential) and heavy-tailed (Pareto) as well as the Cache-follower flow size distribution reported by Facebook \cite{social_inside}. We also considered Poisson flow arrivals with the rate of $\lambda$. We assumed an average flow size of $\mu$ units with a maximum of 500 units along with a minimum size of 2 units for the heavy-tailed distribution.

\vspace{0.5em}
\noindent\textbf{Topologies:} We considered AT\&T North America \cite{att} with 25 nodes and 57 edges, Cogent \cite{cogent} with 197 nodes and 243 edges, GScale \cite{b4} with 12 nodes and 19 edges, AGIS \cite{agis} with 25 nodes and 30 edges, and ANS \cite{ans} with 18 nodes and 25 edges. We assumed bidirectional edges with the capacity of every link in each direction randomly generated using a uniform distribution from 0.2 to 1.

\vspace{0.5em}
\noindent\textbf{Schemes:} We considered three schemes besides \bwrh~and \bwrhf. The \textit{Inverse Capacity Shortest Path} assigns a cost equal to the reciprocal of a link's capacity to it and then selects a shortest hop path from the source to the destination for the new flow. The \textit{Min-Max Utilization} approach chooses a path that has the minimum value of maximum utilization across all paths going from the source to the destination. This approach has been used extensively in the traffic engineering literature \cite{tvlakshman, tempus}. The \textit{Shortest Widest Path} selects the path with the minimum number of hops over which the available bandwidth across the whole path is maximum. The available bandwidth across a path is equal to the minimum of available bandwidth across all its edges.

\vspace{0.5em}
\noindent\textbf{Implementation:} We implemented Algorithm \ref{bwrh} in Java using the JGraphT library. To exhaustively find all paths with at most $K$ hops, we used the class \texttt{AllDirectedPaths} in JGraphT. Algorithm \ref{bwrhf} was also implemented in Java using the class \texttt{DijkstraShortestPath} from JGraphT library.\footnote{We have not made the Java source code publicly available. However, a C++ implementation of \bwrhf~and the exact BWR algorithm can be found in the following Git repository: \url{https://github.com/noormoha/bwr_routing}}


\subsection{Comparison of \bwrh~and \bwrhf}
In Figure \ref{fig:exp_bwrhf_bwrh}, we compare \bwrh~and \bwrhf. We initially compare the mean and tail flow completion times for these two heuristics per topology across various scheduling policies and traffic patterns as shown in Figure \ref{fig:exp_bwrhf_bwrh_diff}. A positive number shows that \bwrh~performed better that \bwrhf~on average. As can be seen, the performance difference is a function of topology. Data points for completion times are averaged across all scenarios per topology. Each scenario was repeated ten times, and for every topology per scenario, the capacity of every link was randomly generated using a uniform distribution from 0.2 to 1. Each sample for computation of the difference was calculated as $\frac{s_{\text{\bwrhf}}-s_{\text{\bwrh}}}{s_{\text{\bwrhf}}}$ where $s$ is either the tail or mean completion times for a simulation instance. Increasing $\mu$ will increase the average network load. The maximum difference in both mean and tail completion times is measured as less than 15\%.

Next, we observe the running time of these two heuristics in Figure \ref{fig:exp_bwrhf_bwrh_running_time}. For the running time, we have reported the worst-case across all runs across all topologies, traffic patterns, and scheduling policies. We can see that the worst-case running time of \bwrhf~remains within the realm of milliseconds while \bwrh~shows a worst-case of tens of seconds. Given that the running time of the routing algorithm is added to the completion time of flows and that these two schemes show little performance difference with regards to completion times of flows, we will focus on \bwrhf~and its further evaluation from here on.

\subsection{Effect of Scheduling Policies}
In Figure \ref{fig:exp_att_policy}, we study the effect of various scheduling policies on the performance of \bwrhf. We see considerable and consistent gains with regards to tail completion times that range from $1.4\times$ to $2.2\times$ across various scheduling policies, traffic patterns, and schemes. We also see significant gains in mean completion times that range from $1.1\times$ to $3.4\times$. Interestingly, we also see that the error bars for \bwrhf~is less than or equal to those of other schemes per metric and per scenario which shows that it offers a more stable performance gain across various scenarios.\footnote{Please note that link capacities are randomly generated per instance ranging from 0.2 to 1 and uniformly distributed. This allows us to examine the schemes over highly varying scenarios given a single connectivity topology.}

\subsection{Effect of Network Topologies}
In Figure \ref{fig:exp_mmf_topology}, we study the effect of various scheduling policies on the performance of \bwrhf. We see that it can improve the mean flow completion times compared to other schemes across different topologies and traffic patterns by up to $1.4\times$. It can also reduce the tail completion times by $1.12\times$ to $2\times$. We also see fewer fluctuations in the performance of \bwrhf, i.e., smaller error bars compared to other schemes. We also see a similar pattern of completion times across different traffic patterns for a given topology, which shows the significant effect of network topology in performance.

\section{Conclusions and Future Directions}
We discussed Best Worst-case Routing (BWR) as an approach to use the remaining volumes of current flows and their paths for improving the completion times of new flows in an online scenario. BWR is a greedy online approach to selecting a path for an incoming flow that optimizes its worst-case completion time. We then noted that this problem is NP-Hard. We also realized that computing the exact worst-case completion time of a new flow on a given path can be computationally intensive due to the interdependence of current flows. We then aimed at developing two heuristics that use two upper bounds on the worst-case completion time for a given path instead of using an exact value. \bwrh~uses a tighter bound on the worst-case completion times but has an exponential worst-case running time, while \bwrhf~uses a looser upper bound with a guaranteed polynomial running time by extending standard shortest path algorithms. We performed extensive evaluations given various topologies, traffic patterns, and scheduling policies and found that \bwrhf~can significantly improve the tail and mean flow completion times. Future directions include the extension of \bwrhf~to multipath routing for further increasing network throughput and a study of how potentially inaccurate flow size information can affect the performance.

{\bibliography{citations.bib}
\bibliographystyle{IEEEtran}}

\begin{figure}[ht]
    \centering
    \subfigure[Computing the worst-case completion time for candidate path $P$]{\includegraphics[width=\columnwidth]{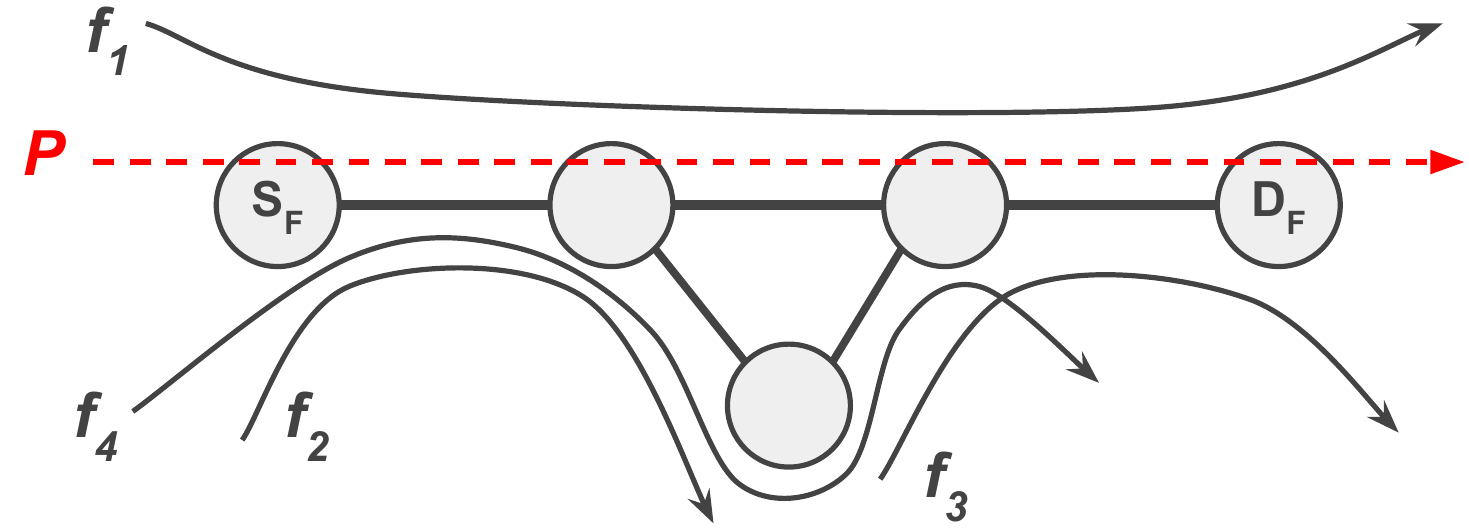}}
    \subfigure[Dependency graph of current flows]{\includegraphics[height=6em]{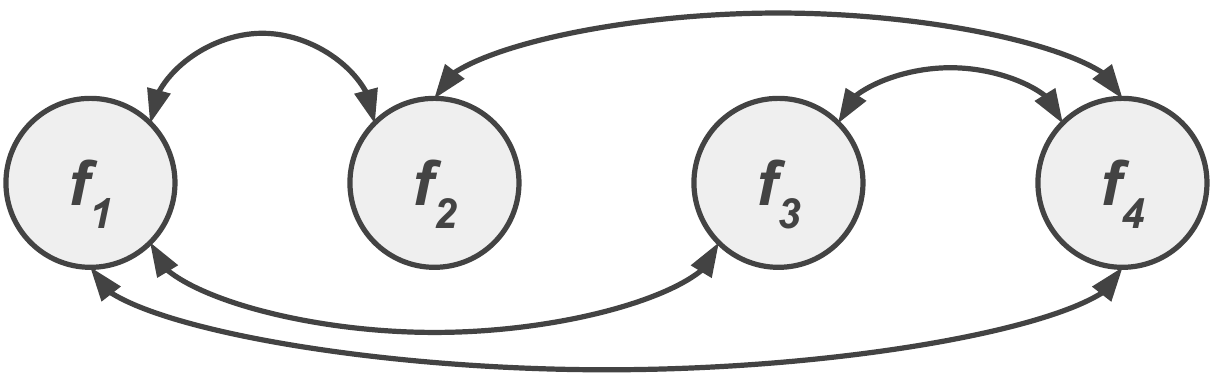}\label{fig:bwr_dep}}
    \caption{A simple scenario with a link capacity of 1.0 for all edges.}
    \label{fig:bwr}
\end{figure}

\appendix
Consider the scenario shown in Figure \ref{fig:bwr}. We want to compute the worst-case completion time of a candidate path $P$ for the new flow $F$ going from $S_F$ to $D_F$ with a volume of $\mathcal{V}_F$, given the four current flows of $f_i, 1 \le i \le 4$ and let us assume that they all have the same remaining volume of $\mathcal{X}$. In Eq. \ref{cost_bwrh}, we computed an upper bound on the worst-case completion time of a new flow by assuming that all the current flows that have a common edge with the new flow will complete sequentially and before the new flow can finish. Therefore, the upper bound will be $4\mathcal{X}+\mathcal{V}_F$ according to Eq. \ref{cost_bwrh}. In this example, we see that it is possible for $f_2$ and $f_3$ to transmit data concurrently. In other words, at any given time, either $f_1$ is transmitting, or $f_4$, or $f_2$ and $f_3$ at the same time. That is, the actual worst-case completion time of $F$ will be $3\mathcal{X}+\mathcal{V}_F$.

In general, to compute the exact worst-case, we will need to use a dependency graph similar to the one in Figure \ref{fig:bwr_dep}. Using the dependency graph, we can figure out which subset of flows can transmit in parallel, that is, any independent set of nodes. In a real scenario with non-uniform link capacities, hundreds of ongoing large flows, and over larger graphs, computing such can be intensive, considering that it has to be repeated per candidate path. Developing a method based on this approach is left to future work.

\end{document}